\documentclass[10pt]{article}
\usepackage{graphicx}
\usepackage{longtable}
\usepackage{times}
\usepackage{setspace}
\usepackage{dcolumn}
\usepackage{bm}
\usepackage{rotating}
\usepackage{hyperref}
\usepackage{epstopdf}
\usepackage{bm}
\usepackage{rotating}
\begin{document}
%\begin{doublespace}
\begin{center}
    \textbf{Gravitational Redshift in Kerr-Newman Geometry Using Gravity's Rainbow}
\end{center}

\begin{center}
    $ Anuj \ Kumar \ Dubey^{1}$,  $ A. \ K. \ Sen^{2}$ and $ Bijoy \ Mazumdar ^{3}$
\end{center}
\begin{center}
    \textit{Department of Physics, Assam University, Silchar-788011, Assam, India}
\end{center}
\begin{center}
    email: $ danuj67@gmail.com^{1}$, $ asokesen@yahoo.com^{2}, bijoymazumdar64@gmail.com^{3}$
\end{center}
\begin{center}
Date: 07 April, 2017
\end{center}
\begin{abstract}
Gravitational redshift is generally reported by most of the authors without considering the influence  of the energy of the test particle using various spacetime geometries such as Schwarzschild, Reissner-Nordstrom, Kerr and Kerr-Newman geometries  for static, charged static, rotating and charged rotating objects respectively. In the present work, the general expression for the energy dependent gravitational redshift is derived for charged rotating body using the Kerr-Newman geometry along with the energy dependent gravity's rainbow function. It is found that the gravitational redshift is influenced by the energy of the source or emitter. One may obtain greater correction in the value of gravitational redshift, using the high energy photons. Knowing the value of gravitational redshift from a high energy sources such as Gamma-ray Bursters (GRB), one may obtain the idea of upper bounds on the dimensionless rainbow function parameter ($\xi$).  Also there may be a possibility to introduce a new physical scale of the order of  $\frac{\xi}{E_{Pl}}$.
\end{abstract}
\section{Introduction}\label{section1}
Gravitational redshift is one of the predictions of General Theory of Relativity. An electromagnetic radiation originating from a source which is in the higher gravitational potential (example pulsars or Sun) is reduced in frequency or redshifted, when observed in a region of a weaker gravitational potential (example earth). Adams in 1925 \cite{Adams} had claimed for the first time the confirmation of the predicted gravitational redshift and Pound and Rebka in 1959 \cite{Pound-Rebka} were the first to experimentally verify the gravitational redshift. Pound and Snider in 1965  \cite{Pound-Snider} had performed an improved version of the experiment of Pound and Rebka to measure the effect of gravity. Snider in 1972 \cite{Snider} had measured the redshift of the solar potassium absorption line. Krisher et al. in 1993 \cite{Krisher} had measured the gravitational redshift of Sun. Nunez and  Nowakowski in 2010 \cite{Nunez} had obtained an expression for gravitational redshift of rotating body by using small perturbations to the Schwarzschild's geometry. Payandeh and Fathi in 2013 \cite{Payandeh} had obtained the gravitational redshift for a static spherically symmetric electrically charged object in Isotropic Reissner-Nordstrom geometry. Dubey and Sen in 2015 \cite{Dubey-Sen-IJTP-2015,Dubey-Sen-ASS-2015} had obtained the expression for gravitational redshift for rotating body and charged rotating body using  Kerr geometry and Kerr-Newman geometry respectively.

The doubly special relativity theory is a modification of special relativity in which a physical energy, which may be the Planck energy remains invariant, in addition to the invariance of speed of light \cite{Magueijo-SmolinPRL2002}. The doubly special relativity is generalized to curved spacetime, and this doubly General Theory of Relativity as proposed by Magueijo and Smolin in 2004 \cite{Magueijo-SmolinCQG2004} is called gravity's rainbow theory. In this theory, the geometry of spacetime depends on the energy of the test particle.

It is mentioned in the work of  Magueijo and Smolin in 2002 \cite{Magueijo-SmolinPRL2002}, there is a possibility of modification for the expression of gravitational redshift to include the energy of the test particle. As stated by Magueijo and Smolin in 2002 \cite{Magueijo-SmolinPRL2002}: \lq \lq The Pound Rebbka experiment is, of course, not sensitive enough for detecting this new effect, but other experiments might be\rq\rq.

With this background the present paper is the extension of the earlier works of Dubey and Sen 2015 \cite{Dubey-Sen-IJTP-2015,Dubey-Sen-ASS-2015}. The present paper is organized as follows:

In Section-\ref{section2}, the modified Kerr-Newman geometry using gravity's rainbow theory is discussed. In Section-\ref{section3}, the  expression for the gravitational redshift in Kerr-Newman geometry is derived. In Section-\ref{section4}, the expression for the gravitational redshift in the modified Kerr-Newman geometry is derived. It is found that the gravitational redshift depends on the energy of the emitted photon. Finally, some conclusions are made in Section-\ref{section5}.

It may be noted that the symbol \lq \ $\tilde{}$ \rq  stands for gravity rainbow effect everywhere in the paper.

\section{Modified spacetime geometry in gravity's rainbow}\label{section2}
 If the mass, charge and rotation (all the three possible characteristics that can be possessed
by a body) is considered, then the geometry of spacetime can be described by using the Kerr-Newman geometry \cite{Newman}. If one considers the three parameters: mass (M), rotation parameter (a) and charge (electric (Q) and / or magnetic (P)), then the most useful form of the solution of Kerr family (Kerr \cite{Kerr} and Kerr-Newman \cite{Newman}) is given in terms of t, r, $\theta$ and $\phi$, where t, and r are Boyer-Lindquist coordinates \cite{Boyer} running from - $\infty$ to + $\infty$, $\theta$  and $\phi$, are ordinary spherical coordinates in which  $\phi$ is periodic with period of 2 $\pi $ and $\theta$ runs from 0 to $\pi$. Covariant form of metric tensor with signature (+,-,-,-) is expressed as:
\begin{equation}\label{1}
ds^{2} = g_{tt}c^{2}dt^{2}+ g_{rr} dr^{2} + g_{\theta\theta} d\theta^{2} +g_{\phi\phi} d\phi^{2} + 2 g_{t \phi} c dt d\phi
\end{equation}
where $g_{ij}$'s are non-zero components of Kerr family.
Non-zero components of $g_{ij}$ of Kerr-Newman metric are given as follows (page 261-262, of Carroll \cite{Carroll}):
\begin{equation}\label{2}
 g_{tt} = (1-\frac{r_{g} r-Q^{2}-P^{2}}{\rho^{2}})
\end{equation}
\begin{equation}\label{3}
g_{rr}=-\frac{\rho^{2}}{\Delta}
\end{equation}
 \begin{equation}\label{4}
 g_{\theta\theta}=-{\rho^{2}}
 \end{equation}
 \begin{equation}\label{5}
g_{\phi\phi}= -[r^{2}+a^{2}+\frac{(r_{g}r-Q^{2}-P^{2}) a^{2}sin^{2}\theta}{\rho^{2}}]sin^{2}\theta
 \end{equation}
\begin{equation}\label{6}
g_{t\phi}=\frac{a sin^{2}\theta (r_{g}r-Q^{2}-P^{2})}{\rho^{2}}
\end{equation}
with
\begin{equation}\label{7}
\rho^{2} = r^{2}+ a^{2}cos^{2}\theta
\end{equation}
and
\begin{equation}\label{8}
 \Delta =r^{2}+ a^{2}-r_{g}r +Q^{2}+P^{2}
\end{equation}
where,
\begin{itemize}
\item $r_g(=2GM/c^2)$ is Schwarzschild radius,
\item M, Q and P are the mass, electric charge and magnetic charge respectively of the central body,
  \item $ a (= \frac{J}{Mc})$ is rotation parameter of the source,
  \item J is the angular momentum of the compact object or central body, which can be also written as $J = I\Omega$,
  \item I is the moment of inertia of the central body,
  \item $\Omega$ is the angular velocity of the central body and
  \item other symbols have their usual meaning.
\end{itemize}
If one puts rotation parameter of the source (a) equal to zero, then the Kerr-Newman metric reduces to Reissner-Nordstrom metric \cite{Reissner,Nordstrom}. Also  if one replaces $(r_{g} r-Q^{2}-P^{2})$ by $(r_{g} r)$ then the Kerr-Newman metric reduces to Kerr metric \cite{Kerr} and further if one puts rotation parameter of the source (a) equal to zero then it reduces to Schwarzschild metric \cite{Schwarzschild}.

Following Dubey and Sen \cite{Dubey-Sen-ASS-2015} the radial components of electric field ($E^{r}$) and magnetic field ($B^{r}$) in terms of electric charge (Q) and magnetic charge (P) can be expressed as:
\begin{equation}\label{9A}
 E^{r}=\frac{Q}{r^{2}+ a^{2}cos^{2}\theta}-\frac{2r(Qr-Pa cos\theta)}{(r^{2}+ a^{2}cos^{2}\theta)^{2}}
\end{equation}
and
$$B^{r}=-[\frac{2}{(r^{2}+ a^{2}cos^{2}\theta) sin\theta}]$$
\begin{equation}\label{9B}
\times[\frac{Qar sin2\theta-P(r^{2}+ a^{2})sin\theta }{r^{2}+ a^{2}cos^{2}\theta}+\frac{(Q a r sin^{2}\theta + P (r^{2}+ a^{2}) cos\theta) a^{2} sin2\theta
 }{(r^{2}+ a^{2}cos^{2}\theta)^{2}}]
\end{equation}
Following Dubey and Sen \cite{Dubey-Sen-IJTP-2015} the four-velocity $u^{i} (=\frac{dx^{i}}{ds})$ of an object in Kerr-Newman geometry can be expressed as,
\begin{equation}\label{10}
u^{i} = (u^{0}, u^{r}, u^{\theta}, u^{\phi})=(u^{0}, 0, 0, u^{0} \ \frac{d\phi}{c dt})=(u^{0}, 0, 0, u^{0} \ \frac{\Omega}{c})
\end{equation}
It may be noted that, on the surface of the central body (example the Sun), $\frac{d\phi}{dt}$ is always the angular velocity $(\Omega$).

The value of the ratio $\frac{u^{\phi}}{u^{0}}$, can be written as:
\begin{equation}\label{10A}
\frac{u^{\phi}}{u^{0}} = \frac{\Omega}{c}
\end{equation}
 Here the time component of the four velocity $(u^{0})$ is expressed as:
\begin{equation}\label{11}
 u^{0}=\frac{1}{\sqrt{g_{tt}+ g_{\phi\phi}(\frac{\Omega}{c})^{2} +2 g_{t \phi}(\frac{\Omega}{c})}}
\end{equation}

This is not difficult to obtain the modified spacetime metric in gravity's rainbow theory.

Following certain analogy (as discussed by Magueijo and Smolin in 2004 \cite{Magueijo-SmolinCQG2004}), one can simply replace time coordinate $dx^{0}$ by $\frac{dx^{0}}{f(\frac{E}{E_{Pl}})}$ and all the spatial coordinates $dx^{i}$ by $\frac{dx^{i}}{g(\frac{E}{E_{Pl}})}$ to obtain the modified spacetime metric in gravity's rainbow theory.

Here E and $E_{Pl}$ are the particle energy and the Planck energy respectively. The functions $f(\frac{E}{E_{Pl}}$) and $g(\frac{E}{E_{Pl}}$) are called the rainbow functions. The main property of the gravity's rainbow functions is that they make spacetime energy dependent and this is the main requirement of the present work for deriving the energy dependent expression of gravitational redshift.

 Now the modified Kerr-Newman spacetime metric $d\tilde{s}^{2}$ in gravity's rainbow theory can be expressed as:
\begin{equation}\label{12}
d\tilde{s}^{2} = g_{tt}\frac{c^{2} dt^{2}}{f^{2}(\frac{E}{E_{Pl}})}+ g_{rr} \frac{dr^{2}}{g^{2}(\frac{E}{E_{Pl}})} + g_{\theta\theta} \frac{d\theta^{2}}{g^{2}(\frac{E}{E_{Pl}})} +g_{\phi\phi} \frac{d\phi^{2}}{g^{2}(\frac{E}{E_{Pl}})} + 2 g_{t \phi}  \frac{cdt}{f(\frac{E}{E_{Pl}})} \frac{d\phi}{g(\frac{E}{E_{Pl}})}
\end{equation}
The four-velocity $\tilde{u}^{i} (=\frac{dx^{i}}{d\tilde{s}})$ of an object in the modified Kerr-Newman geometry can be expressed as,
\begin{equation}\label{13A}
\tilde{u}^{i} = (\tilde{u}^{0}, \tilde{u}^{r}, \tilde{u}^{\theta}, \tilde{u}^{\phi})=(\tilde{u}^{0}, 0, 0, \tilde{u}^{0} \ \frac{\Omega}{c})
\end{equation}
The value of the ratio $\frac{\tilde{u}^{\phi}}{\tilde{u}^{0}}$, can be written as:
\begin{equation}\label{13B}
\frac{\tilde{u}^{\phi}}{\tilde{u}^{0}} =  \frac{\Omega}{c}
\end{equation}
In the modified Kerr-Newman spacetime the time component of the four velocity $(\tilde{u}^{0})$ can be expressed as:
\begin{equation}\label{13C}
 \tilde{u}^{0}=\frac{1}{\sqrt{\frac{g_{tt}}{f^{2}(\frac{E}{E_{Pl}})}+ \frac{g_{\phi\phi}}{g^{2}(\frac{E}{E_{Pl}})}(\frac{\Omega}{c})^{2} +2 \frac{g_{t\phi}}{f(\frac{E}{E_{Pl}}) \ g(\frac{E}{E_{Pl}})}(\frac{\Omega}{c})}}
\end{equation}
In gravity's rainbow theory, the modified energy-momentum dispersion relation is discussed by various authors (as \cite{Hendi-FaizalPRD2015}). In gravity's rainbow theory, the modified energy-momentum dispersion relation is given as:
\begin{equation}\label{14}
    E^{2}f^{2} (\frac{E}{E_{Pl}})-p^{2} c^{2} g^{2} (\frac{E}{E_{Pl}})= m^{2} c^{4}
\end{equation}
The gravity's rainbow functions are required to satisfy the conditions:
$\lim _{\frac{E}{E_{Pl}}\rightarrow 0} f(\frac{E}{E_{Pl}})=1$ and
$\lim _{\frac{E}{E_{Pl}}\rightarrow 0} g(\frac{E}{E_{Pl}})=1$.
This condition is required, as the theory is constrained to reproduce the standard dispersion relation in the smaller energy limit.

\section{Gravitational Redshift in Kerr-Newman Geometry}\label{section3}

From the expression $ds^{2}=c^{2} d\tau^{2}=g_{ik}dx^{i}dx^{k}$, this is clear that, the rate of clock is influenced by the gravitational field.  One needs to compare data from two different points in a gravitational field, in order to find an observable effect. One could consider two observers, a locally inertial observer at the emitter or source and an asymptotic observer. A locally inertial observer  measures the frequency $\omega$, which may be considered as the frequency at the location of source (emitter). While an asymptotic observer measures the frequency $\omega^{'}$, which may be considered as the frequency at the location of observer.

In General Theory of Relativity gravitational redshift (Z) and gravitational redshift factor $(\Re)$ are defined as \cite{Pineault-Roeder,Cunningham,Muller,Dubey-Sen-IJTP-2015,Dubey-Sen-ASS-2015}:
\begin{equation}\label{GR Redshift}
Z=\frac{1}{\Re}-1=\frac{\omega^{'}}{\omega}-1
\end{equation}

Following (page 141, of Landau and Lifshitz \cite{Landau}), the wave four-vector $k_{i}$ can be written as:
\begin{equation}\label{3.4}
k_{i}=-\frac{\partial\Psi}{\partial x^{i}}
\end{equation}
where $\Psi$, is defined as eikonal.

Using the above equation (\ref{3.4}), the components of wave four-vector ($k_{i}$) are given as:
\begin{equation}\label{3.7}
k_{0}=-\frac{\partial\Psi}{\partial x^{0}}= -\frac{\partial\Psi}{\partial ct}=\frac{\omega^{'}}{c}
\end{equation}
\begin{equation}\label{3.8}
k_{r}= -\frac{\partial\Psi}{\partial r}
\end{equation}
\begin{equation}\label{3.9}
k_{\theta}=-\frac{\partial\Psi}{\partial \theta}
\end{equation}
\begin{equation}\label{3.10}
k_{\phi}= -\frac{\partial\Psi}{\partial \phi}
\end{equation}

Frequency $\omega$ measured in terms of proper time ($\tau$) is defined as (page 268, of Landau and Lifshitz \cite{Landau}):
\begin{equation}\label{3.12}
\omega=-\frac{\partial\Psi}{\partial \tau}
\end{equation}

It should be noted that the frequency $\omega$  expressed in terms of proper time ($\tau$), is different at different point of space. While the frequency $\omega^{'}$ expressed in terms of asymptotic observers's time (t) is constant (page 268, of Landau and Lifshitz \cite{Landau}).

 The above equation (\ref{3.12}) can be written as:
\begin{equation}\label{3.13}
\frac{\omega}{c}=-\frac{\partial\Psi}{\partial c \tau}=-(\frac{\partial\Psi}{\partial c t}) \ (\frac{\partial c t}{\partial c\tau})- (\frac{\partial\Psi}{\partial r}) \ (\frac{\partial r}{ \partial c \tau})- (\frac{\partial\Psi}{\partial \theta}) \ (\frac{\partial \theta}{ \partial c \tau})- (\frac{\partial\Psi}{\partial \phi}) \ (\frac{\partial \phi}{\partial c \tau})
\end{equation}
Now using the values of $u^{i} (=\frac{dx^{i}}{ds})$ and also using the value of $k_{i}(=-\frac{\partial\Psi}{\partial x^{i}})$ from equations (\ref{3.7}) to (\ref{3.10}), in the above equation (\ref{3.13}), it becomes,
\begin{equation}\label{3.14}
\frac{\omega}{c}= k_{0} \ u^{0} + k_{r} \ u^{r} + k_{\theta} \ u^{\theta} + k_{\phi} \ u^{\phi}=k_{i} \ u^{i}
\end{equation}
For a sphere, the photon is emitted at a location on its surface where $dr = d\theta =0$, as the sphere rotates. Now the above equation (\ref{3.14}) can be written as:
\begin{equation}\label{3.15A}
[k_{i} \ u^{i}]_{Source} = [k_{0} \ u^{0} + k_{\phi} \ u^{\phi}]_{Source}= [k_{0} \ u^{0}(1 + \frac{k_{\phi}}{k_{0}} \ \frac{u^{\phi}}{u^{0}})]_{Source}
\end{equation}
and
\begin{equation}\label{3.15B}
[k_{i} \ u^{i}]_{Observer} = [k_{0} \ u^{0} + k_{\phi} \ u^{\phi}]_{Observer}= [k_{0} \ u^{0}(1 + \frac{k_{\phi}}{k_{0}} \ \frac{u^{\phi}}{u^{0}})]_{Observer}
\end{equation}
For asymptotic observer,
\begin{equation}\label{3.15C}
  u^{0}_{Observer}=1
\end{equation}
 and
\begin{equation}\label{3.15D}
  u^{\phi}_{Observer}=0
\end{equation}
One knows $k_{i} \ u^{i}$ is an invariant. Therefore
\begin{equation}\label{New1}
  [k_{i} \ u^{i}]_{Source} = [k_{i} \ u^{i}]_{Observer}
\end{equation}
and
\begin{equation}\label{New2}
  [k_{0} \ u^{0}(1 + \frac{k_{\phi}}{k_{0}} \ \frac{u^{\phi}}{u^{0}} )]_{Source} = [k_{0}]_{Observer}
\end{equation}
Here $u^{0}_{Source}(=\frac{1}{\sqrt{g_{tt}+ g_{\phi\phi}(\frac{\Omega}{c})^{2} +2 g_{t \phi}(\frac{\Omega}{c})}})$ is the time component of the four-velocity in Kerr-Newman geometry as given by equation (\ref{11}). The value of $\frac{u^{\phi}_{Source}}{u^{0}_{Source}}(=\frac{\Omega}{c})$ is given by equation (\ref{10A}).

Substituting the values of $[k_{0}]_{Source}(=\frac{\omega}{c})$ and $[k_{0}]_{Observer}(=\frac{\omega^{'}}{c})$ in the above equation (\ref{New2}), it becomes,
\begin{equation}\label{New3}
 \frac{\omega}{c} \  [\frac{1 + \frac{k_{\phi}}{k_{0}} \ \frac{\Omega}{c}}{\sqrt{g_{tt}+ g_{\phi\phi}(\frac{\Omega}{c})^{2} +2 g_{t \phi}(\frac{\Omega}{c})}}]  = \frac{\omega^{'}}{c}
\end{equation}
At the location of the observer the observed frequency is defined by $\omega^{'}$.

The above equation (\ref{New3}) can be written as:
\begin{equation}\label{extra}
\omega^{'}=  [\frac{\omega}{\sqrt{g_{tt}+ g_{\phi\phi}(\frac{\Omega}{c})^{2} +2 g_{t \phi}(\frac{\Omega}{c})}}+\frac{ \omega \ (\frac{k_{\phi}}{k_{0}} \ \frac{\Omega}{c})}{\sqrt{g_{tt}+ g_{\phi\phi}(\frac{\Omega}{c})^{2} +2 g_{t \phi}(\frac{\Omega}{c})}}]
\end{equation}

To proceed with further calculations, now one needs the value of $[\frac{k_{\phi}}{k_{0}}]_{Source}$.\\

In the gravitational field of a rotating spherical mass, the relativistic action function \lq S\rq  \ for a particle with the coordinate time \lq t\rq  \ and the angle \lq $\phi$\rq  \  as cyclic variables, can be expressed as (page 264, 328 of Landau and Lifshitz \cite{Landau}), Eq. (13) of Roy and Sen \cite{Roy}):

\begin{equation}\label{3.16}
S = - E t + L \phi + S_{r} (r)+ S_{\theta} (\theta)
\end{equation}
where \lq E\rq \ is the conserved energy and \lq L\rq \ denotes the component of the angular momentum along the axis of the symmetry of the field.

The four-momentum ($p_{i}$) is defined as (page 264, of Landau and Lifshitz \cite{Landau})):

\begin{equation}\label{3.17}
p_{i}=-\frac{\partial S}{\partial x^{i}}
\end{equation}

Using the above equations (\ref{3.16}) and (\ref{3.17}), one can write:

\begin{equation}\label{3.18}
p_{t}=-\frac{\partial S}{\partial x^{0}}=\frac{E}{c}
\end{equation}
and
\begin{equation}\label{3.19}
p_{\phi}=-\frac{\partial S}{\partial \phi}=-L
\end{equation}

Now for photon propagation, replacing the action (S) by the eikonal $(\Psi)$ and following certain analogies (page 264 and 265, of Landau and Lifshitz \cite{Landau})), the above equations (\ref{3.18}) and (\ref{3.19}) becomes,

\begin{equation}\label{3.20}
k_{0}=-\frac{\partial \Psi}{\partial x^{0}}=\frac{E}{c}
\end{equation}
and
\begin{equation}\label{3.21}
k_{\phi}=-\frac{\partial \Psi}{\partial \phi}=-L
\end{equation}

Now using the values of $k_{0}$ and $k_{\phi}$ from the above equations (\ref{3.20}) and (\ref{3.21}), the value of $\frac{k_{\phi}}{k_{0}}$  can be written as:
\begin{equation}\label{3.22}
\frac{k_{\phi}}{k_{0}} = \frac{L \ c}{E}
\end{equation}

In case of photon, the energy (E) and linear momentum (p) are expressed as:
\begin{equation}\label{2.41}
E^{2} = p^{2} \ c^{2}+m_{0}^{2} \ c^{4}
\end{equation}
where $m_{0}$ is rest mass of the photon which is zero. So above equation (\ref{2.41}) becomes:
\begin{equation}\label{2.42}
E = p \ c
\end{equation}
Considering the ray of light which is emitted radially outward from a point $(r,\theta,\phi)$ at the rotating object (for example may be, a pulsar or Sun). The angular momentum (L) about the symmetry axis may be expressed as:
\begin{equation}\label{2.43}
L = p \ r \ sin\phi \  sin\theta
\end{equation}

It may be noted that, in the above equation (\ref{2.43}), the value of angular momentum \lq L\rq \ will become zero as $\phi$ is zero. The \lq $sin\phi$\rq \ term was introduced, in order to capture the motion of emitter along the line of sight as measured from the location of the observer.

Substituting the values of E and L from the above equations (\ref{2.42}) and (\ref{2.43}) respectively, in equation (\ref{3.22}), the values of $\frac{k_{\phi}}{k_{0}}$ from equation (\ref{3.22}) can be rewritten as:
\begin{equation}\label{3.23}
\frac{k_{\phi}}{k_{0}} = \frac{p \ r \ sin\phi \  sin\theta \ c}{p \ c}=  \ r \ sin\phi \  sin\theta
\end{equation}
Substituting the value of $\frac{k_{\phi}}{k_{0}}$ from the above equation (\ref{3.23}) in equation (\ref{extra}), it becomes,
\begin{equation}{\label{3.24}}
\omega^{'} = \frac{\omega}{\sqrt{g_{tt}+ g_{\phi\phi}(\frac{\Omega}{c})^{2} +2 g_{t \phi}(\frac{\Omega}{c})}} +\frac{\omega\ r \ sin\phi \  sin\theta\ \frac{\Omega}{c}}{\sqrt{g_{tt}+ g_{\phi\phi}(\frac{\Omega}{c})^{2} +2 g_{t \phi}(\frac{\Omega}{c})}}
\end{equation}

Now using the expressions (\ref{GR Redshift}) and (\ref{3.24}), one can write the expression of gravitational redshift $(Z_{K-N})$ in Kerr-Newman geometry as:
\begin{equation}{\label{3.250}}
Z_{K-N}=[\frac{1 + \ r \ sin\phi \  sin\theta \ \frac{\Omega}{c}}{\sqrt{g_{tt}+ g_{\phi\phi}(\frac{\Omega}{c})^{2} +2 g_{t \phi}(\frac{\Omega}{c})}} ]-1
\end{equation}
Substituting the values of metric coefficients $g_{tt}$, $g_{\phi\phi}$, and $g_{t\phi}$ from equations (\ref{2}), (\ref{5}) and (\ref{6}) of Kerr-Newman metric in the above expression (\ref{3.250}), now one can write the expression for gravitational redshift ($Z_{K-N}(\phi,\theta)$) in Kerr-Newman geometry for a charged rotating body with electric charge (Q) and magnetic charge (P) as:
$$Z_{K-N}(\phi,\theta)=$$
\begin{equation}\label{4.29}
 [\frac{(1 + \ r \ sin\phi \  sin\theta \ \frac{\Omega}{c})}{\sqrt{(1-\frac{r_{g} r-Q^{2}-P^{2}}{r^{2}+ a^{2}cos^{2}\theta}) -[(r^{2}+a^{2}+\frac{(r_{g} r-Q^{2}-P^{2}) a^{2}sin^{2}\theta}{r^{2}+ a^{2}cos^{2}\theta})sin^{2}\theta] (\frac{\Omega}{c})^{2} +2 (\frac{a sin^{2}\theta (r_{g}r-Q^{2}-P^{2})}{r^{2}+ a^{2}cos^{2}\theta})(\frac{\Omega}{c})}}]-1
\end{equation}

Substituting the value of electric charge (Q) and magnetic charge (P) both equals to zero in the above expression (\ref{4.29}), the above expression (\ref{4.29}) becomes,
$$Z_{Kerr}(\phi,\theta)=$$
\begin{equation}{\label{3.25}}
[\frac{(1 + \ r \ sin\phi \  sin\theta \ \frac{\Omega}{c})}{\sqrt{(1-\frac{r_{g} r}{r^{2}+ a^{2}cos^{2}\theta}) -[(r^{2}+a^{2}+\frac{r_{g} r a^{2}sin^{2}\theta}{r^{2}+ a^{2}cos^{2}\theta})sin^{2}\theta] (\frac{\Omega}{c})^{2} +2 (\frac{a sin^{2}\theta r_{g}r}{r^{2}+ a^{2}cos^{2}\theta})(\frac{\Omega}{c})}}]-1
\end{equation}
The above expression (\ref{3.25}) is the general expression of gravitational redshift $(Z_{Kerr}(\phi,\theta))$ in Kerr geometry. Using this expression (\ref{3.25}), one can obtain the value of gravitational redshift (Z) for the values of $\theta$ from zero (pole) to $\frac{\pi}{2}$ (equator) and $\phi$ from zero to $2\pi$.

Thus the redshift as given by the above expression (\ref{3.25}) clearly depends on the coordinates $\theta$ and $\phi$.

Considering the ray of light which is emitted from a point on the equatorial plane $(r,\theta=\frac{\pi}{2},\phi)$ at the rotating object, the equation (\ref{3.25}) of gravitational redshift $(Z_{Kerr}(\phi,\theta))$ becomes,
\begin{equation}\label{3.26}
Z_{Kerr}(\phi,\theta=\frac{\pi}{2})= [\frac{1 + r \ sin\phi \ \frac{\Omega}{c}}{\sqrt{(1-\frac{r_{g}}{r})- (r^{2}+a^{2}+\frac{r_{g}a^{2}}{r})(\frac{\Omega}{c})^{2} +2 (\frac{r_{g}a}{r})(\frac{\Omega}{c})}}]-1
\end{equation}
Using the above expression (\ref{3.26}) of gravitational redshift $Z(\phi,\theta=\frac{\pi}{2})$, one can obtain the variation of gravitational redshift (Z) with respect to azimuthal coordinate $(\phi)$ from zero to $2\pi$.

At fixed value of $\phi=\frac{\pi}{2}$, the expression of gravitational redshift (\ref{3.25}) becomes,
$$Z_{Kerr}(\phi=\frac{\pi}{2},\theta)= $$
\begin{equation}\label{3.27}
[\frac{(1 + \ r \ sin\theta \ \frac{\Omega}{c})}{\sqrt{(1-\frac{r_{g} r}{r^{2}+ a^{2}cos^{2}\theta}) -[(r^{2}+a^{2}+\frac{r_{g} r a^{2}sin^{2}\theta}{r^{2}+ a^{2}cos^{2}\theta})sin^{2}\theta] (\frac{\Omega}{c})^{2} +2 (\frac{a sin^{2}\theta r_{g}r}{r^{2}+ a^{2}cos^{2}\theta})(\frac{\Omega}{c})}}]-1
\end{equation}
Using the above expression (\ref{3.27}) of gravitational redshift $(Z_{Kerr}(\phi=\frac{\pi}{2},\theta))$, one can obtain the gravitational redshift for the values of $\theta$ from zero (pole) to $\frac{\pi}{2}$ (equator). This shows clearly the dependence of gravitational redshift on the latitude.

In the expression (\ref{4.29}) of the gravitational redshift $(Z_{K-N}(\phi,\theta))$,  if one substitutes magnetic charge (P) and the rotation parameter (a) both equal to zero, then one can obtain the corresponding redshift in  Reissner-Nordstrom geometry. The expression of gravitational redshift $(Z_{R-N})$ in Reissner-Nordstrom geometry can be written as:

\begin{equation}\label{3.28A}
Z_{R-N}=\frac{1}{\sqrt{1-\frac{r_{g}}{r}-\frac{Q^{2}} {r^{2}}}}-1
\end{equation}
Further, if one considers the electric charge (Q) to be zero in the above equation (\ref{3.28A}) of gravitational redshift $(Z_{R-N})$ in Reissner-Nordstrom geometry, then it becomes,

\begin{equation}\label{3.28}
Z_{Schwrzschild}=\frac{1}{\sqrt{1-\frac{r_{g}}{r}}}-1
\end{equation}
The above equation (\ref{3.28}) is the expression of corresponding gravitational redshift from a static body of same mass (Schwarzschild mass). In  most of the published works the similar expression is given for the gravitational redshift $(Z_{Schwrzschild})$ for radiation from the stellar surface in the spherically symmetric (Schwarzschild) gravitational field.

Now as the photon is emitted from the surface of the star (pulsar or Sun), one can substitute the actual value for \lq r\rq, which is the physical radius of the star \lq R\rq \ for carrying out some numerical calculations. Thus, the information on the ratio $ (\frac{M}{R})$ for a compact object can be obtained from the gravitational redshift of a static body, where M and R are mass and the physical radius respectively of the central body. And, in case of the Kerr-Newman body, one can get additional information about rotation velocity and the charge.

\section{Gravitational Redshift in Modified Kerr-Newman Geometry}\label{section4}
The main aim of the present work is to obtain the energy dependent expression for gravitational redshift in modified Kerr-Newman geometry using the gravity's rainbow theory. This can be achieved simply by replacing the non-zero components $g_{ij}$ of Kerr-Newman metric $g_{tt}$ by $\frac{g_{tt}}{f^{2}(\frac{E}{E_{Pl}})}$, $g_{\phi\phi}$ by $\frac{g_{\phi\phi}}{g^{2}(\frac{E}{E_{Pl}})}$ and $g_{t\phi}$ by $\frac{g_{t\phi}}{f(\frac{E}{E_{Pl}}) \ g(\frac{E}{E_{Pl}})}$.

After making the above stated replacement in the expression (\ref{3.250}) or (\ref{4.29}), one can write the expression for gravitational redshift ($\tilde{Z}_{K-N}(\phi,\theta)$) in the modified Kerr-Newman geometry for a charged rotating body with electric charge (Q) and magnetic charge (P) as:
$$\tilde{Z}_{K-N}(\phi,\theta)=$$
\begin{equation}\label{K-N}
 [\frac{(1 + \ r \ sin\phi \  sin\theta \ \frac{\Omega}{c})}{\sqrt{\frac{(1-\frac{r_{g} r-Q^{2}-P^{2}}{r^{2}+ a^{2}cos^{2}\theta})}{f^{2}(\frac{E}{E_{Pl}})}- \frac{[(r^{2}+a^{2}+\frac{(r_{g} r-Q^{2}-P^{2}) a^{2}sin^{2}\theta}{r^{2}+ a^{2}cos^{2}\theta})sin^{2}\theta]}{g^{2}(\frac{E}{E_{Pl}})}(\frac{\Omega}{c})^{2} +2 \frac{(\frac{a sin^{2}\theta (r_{g}r-Q^{2}-P^{2})}{r^{2}+ a^{2}cos^{2}\theta})}{f(\frac{E}{E_{Pl}}) \ g(\frac{E}{E_{Pl}})}(\frac{\Omega}{c})}}]-1
\end{equation}
It should be noted that, the appropriate selection of the rainbow functions are very important for making the predictions. This selection is preferred to be based on the phenomenological motivations.

Among the various proposals for the choice of rainbow functions, one can take the following rainbow function as proposed by Magueijo and Smolin in 2004 \cite{Magueijo-SmolinCQG2004}:
\begin{equation}\label{Function}
f(\frac{E}{E_{Pl}})=g(\frac{E}{E_{Pl}})=\frac{1}{1-\xi\frac{E}{E_{Pl}}}
\end{equation}
This particular choice of the rainbow function does not provide a varying speed of light and yet might solve the horizon problem \cite{Magueijo-SmolinCQG2004}.

Here $\xi$ is some dimensionless parameter and generally assumed that this parameter $\xi$ is of the order of unity. Ali and Khalil in 2015 \cite{Ali-KhalilEPL2015} had discussed that, the upper bound on  this parameter of the rainbow function ($\xi$) is $\xi<8.5\times10^{21}$.

Substituting the value of rainbow functions $f(\frac{E}{E_{Pl}})=g(\frac{E}{E_{Pl}})$ from equation (\ref{Function}) in the above equation (\ref{K-N}), then the equation (\ref{K-N}) becomes,
$$\tilde{Z}_{K-N}(\phi,\theta)=$$
\begin{equation}\label{K-N1}
 [\frac{(1 + \ r \ sin\phi \  sin\theta \ \frac{\Omega}{c}) \ (\frac{1}{1-\xi\frac{E}{E_{Pl}}})}{\sqrt{(1-\frac{r_{g} r-Q^{2}-P^{2}}{r^{2}+ a^{2}cos^{2}\theta}) -[(r^{2}+a^{2}+\frac{(r_{g} r-Q^{2}-P^{2}) a^{2}sin^{2}\theta}{r^{2}+ a^{2}cos^{2}\theta})sin^{2}\theta] (\frac{\Omega}{c})^{2} +2 (\frac{a sin^{2}\theta (r_{g}r-Q^{2}-P^{2})}{r^{2}+ a^{2}cos^{2}\theta})(\frac{\Omega}{c})}}]-1
\end{equation}
After simplification (using binomial approximation) and neglecting the second and higher order terms of $(\frac{E}{E_{Pl}})$, the above equation (\ref{K-N1}) can be written as:
$$\tilde{Z}_{K-N}(\phi,\theta)=$$
\begin{equation}\label{K-N2}
 [\frac{(1 + \ r \ sin\phi \  sin\theta \ \frac{\Omega}{c}) \ (1+\xi\frac{E}{E_{Pl}})}{\sqrt{(1-\frac{r_{g} r-Q^{2}-P^{2}}{r^{2}+ a^{2}cos^{2}\theta}) -[(r^{2}+a^{2}+\frac{(r_{g} r-Q^{2}-P^{2}) a^{2}sin^{2}\theta}{r^{2}+ a^{2}cos^{2}\theta})sin^{2}\theta] (\frac{\Omega}{c})^{2} +2 (\frac{a sin^{2}\theta (r_{g}r-Q^{2}-P^{2})}{r^{2}+ a^{2}cos^{2}\theta})(\frac{\Omega}{c})}}]-1
\end{equation}
Considering the angular frequency of the emitted photon as $\omega^{'}_{S}$, then the energy of the emitted photon (E) becomes $\hbar \omega^{'}_{S}$. Now the above equation (\ref{K-N2}) can be rewritten as:
$$\tilde{Z}_{K-N}(\phi,\theta)=$$

 $$[\frac{1}{\sqrt{(1-\frac{r_{g} r-Q^{2}-P^{2}}{r^{2}+ a^{2}cos^{2}\theta}) -[(r^{2}+a^{2}+\frac{(r_{g} r-Q^{2}-P^{2}) a^{2}sin^{2}\theta}{r^{2}+ a^{2}cos^{2}\theta})sin^{2}\theta] (\frac{\Omega}{c})^{2} +2 (\frac{a sin^{2}\theta (r_{g}r-Q^{2}-P^{2})}{r^{2}+ a^{2}cos^{2}\theta})(\frac{\Omega}{c})}}-1]$$

 $$+[\frac{( r \ sin\phi \  sin\theta \ \frac{\Omega}{c})}{\sqrt{(1-\frac{r_{g} r-Q^{2}-P^{2}}{r^{2}+ a^{2}cos^{2}\theta}) -[(r^{2}+a^{2}+\frac{(r_{g} r-Q^{2}-P^{2}) a^{2}sin^{2}\theta}{r^{2}+ a^{2}cos^{2}\theta})sin^{2}\theta] (\frac{\Omega}{c})^{2} +2 (\frac{a sin^{2}\theta (r_{g}r-Q^{2}-P^{2})}{r^{2}+ a^{2}cos^{2}\theta})(\frac{\Omega}{c})}}]$$
 \begin{equation}\label{K-N3}
+ [\frac{(\xi\frac{\hbar \omega^{'}_{S}}{E_{Pl}}) \ (1 + \ r \ sin\phi \  sin\theta \ \frac{\Omega}{c})}{\sqrt{(1-\frac{r_{g} r-Q^{2}-P^{2}}{r^{2}+ a^{2}cos^{2}\theta}) -[(r^{2}+a^{2}+\frac{(r_{g} r-Q^{2}-P^{2}) a^{2}sin^{2}\theta}{r^{2}+ a^{2}cos^{2}\theta})sin^{2}\theta] (\frac{\Omega}{c})^{2} +2 (\frac{a sin^{2}\theta (r_{g}r-Q^{2}-P^{2})}{r^{2}+ a^{2}cos^{2}\theta})(\frac{\Omega}{c})}}]
\end{equation}
In the above equation (\ref{K-N3}), the first term is purely due to time dilation, which is reported by most of the authors, as gravitational redshift, the second term is  due to the rotation of the emitter of photon in a Kerr-Newman field, which is Doppler like and the additional third term is due to the frequency of the emitter itself.

The above expression (\ref{K-N3}) for gravitational redshift ($\tilde{Z}_{K-N}(\phi,\theta)$) in the modified Kerr-Newman geometry for a charged rotating body with electric charge (Q) and magnetic charge (P) can also be written as:

\begin{equation}\label{K-N4}
\tilde{Z}_{K-N}(\phi,\theta)=Z_{K-N}(\phi,\theta) + \xi\frac{\hbar \omega^{'}_{S}}{E_{Pl}} \ (Z_{K-N}(\phi,\theta)+1)
\end{equation}
In the limit \ $\frac{\hbar \omega^{'}_{S}}{E_{Pl}}\longrightarrow0$, the above equation becomes,
\begin{equation}\label{K-N5}
\tilde{Z}_{K-N}(\phi,\theta)=Z_{K-N}(\phi,\theta)
\end{equation}

If one substitutes electric charge (Q) and magnetic charge (P) both equal to zero, in the expressions (\ref{K-N3}) and (or) (\ref{K-N4}) of the modified gravitational redshift $(\tilde{Z}_{K-N}(\phi,\theta))$, then one can obtain the corresponding redshift in Kerr geometry. The modified expression of gravitation redshift $(\tilde{Z}_{Kerr}(\phi,\theta))$ in Kerr geometry can be written as:

\begin{equation}\label{Kerr}
\tilde{Z}_{Kerr}(\phi,\theta)=Z_{Kerr}(\phi,\theta) + \xi\frac{\hbar \omega^{'}_{S}}{E_{Pl}} \ (Z_{Kerr}(\phi,\theta)+1)
\end{equation}
In the expression (\ref{K-N3}) of the modified gravitational redshift $(\tilde{Z}_{K-N}(\phi,\theta))$,  if one substitutes magnetic charge (P) and the rotation parameter (a) or angular velocity $\Omega$ both equal to zero, then one can obtain the corresponding redshift in  Reissner-Nordstrom geometry. The modified expression of gravitation redshift $(\tilde{Z}_{R-N})$ in Reissner-Nordstrom geometry can be written as:
\begin{equation}\label{R-N}
\tilde{Z}_{R-N}= Z_{R-N}(\phi,\theta) + \xi\frac{\hbar \omega^{'}_{S}}{E_{Pl}} \ (Z_{R-N}(\phi,\theta)+1)
\end{equation}
Further, if one substitutes the electric charge (Q) equal to zero in the above equation (\ref{R-N}), then one can obtain the corresponding redshift in Schwarzschild geometry. The modified expression of gravitation redshift $(\tilde{Z}_{Schwarzschild})$ in Schwarzschild geometry can be written as:
\begin{equation}\label{Schwarzschild}
\tilde{Z}_{Schwarzschild}= Z_{Schwarzschild} + \xi\frac{\hbar \omega^{'}_{S}}{E_{Pl}} \ (Z_{Schwarzschild}
+1)
\end{equation}
Now the obtained expression of gravitational redshift (\ref{Schwarzschild}) is
matches with the gravitational redshift for a static spherically symmetric object in the modified Schwarzschild geometry (Eqn. (34) of Ali and Khalil 2015 \cite{Ali-KhalilEPL2015}).

The presence of an additional term ($\xi\frac{\hbar \omega^{'}_{S}}{E_{Pl}}$) in the expressions (\ref{K-N4}), (\ref{Kerr}), (\ref{R-N}) and (\ref{Schwarzschild}) for the gravitational redshift, clearly indicates that, one can obtain greater correction in the value of gravitational redshift, using the high energy photons.

\begin{figure*}
\includegraphics[width=40pc, height=30pc,angle=270]{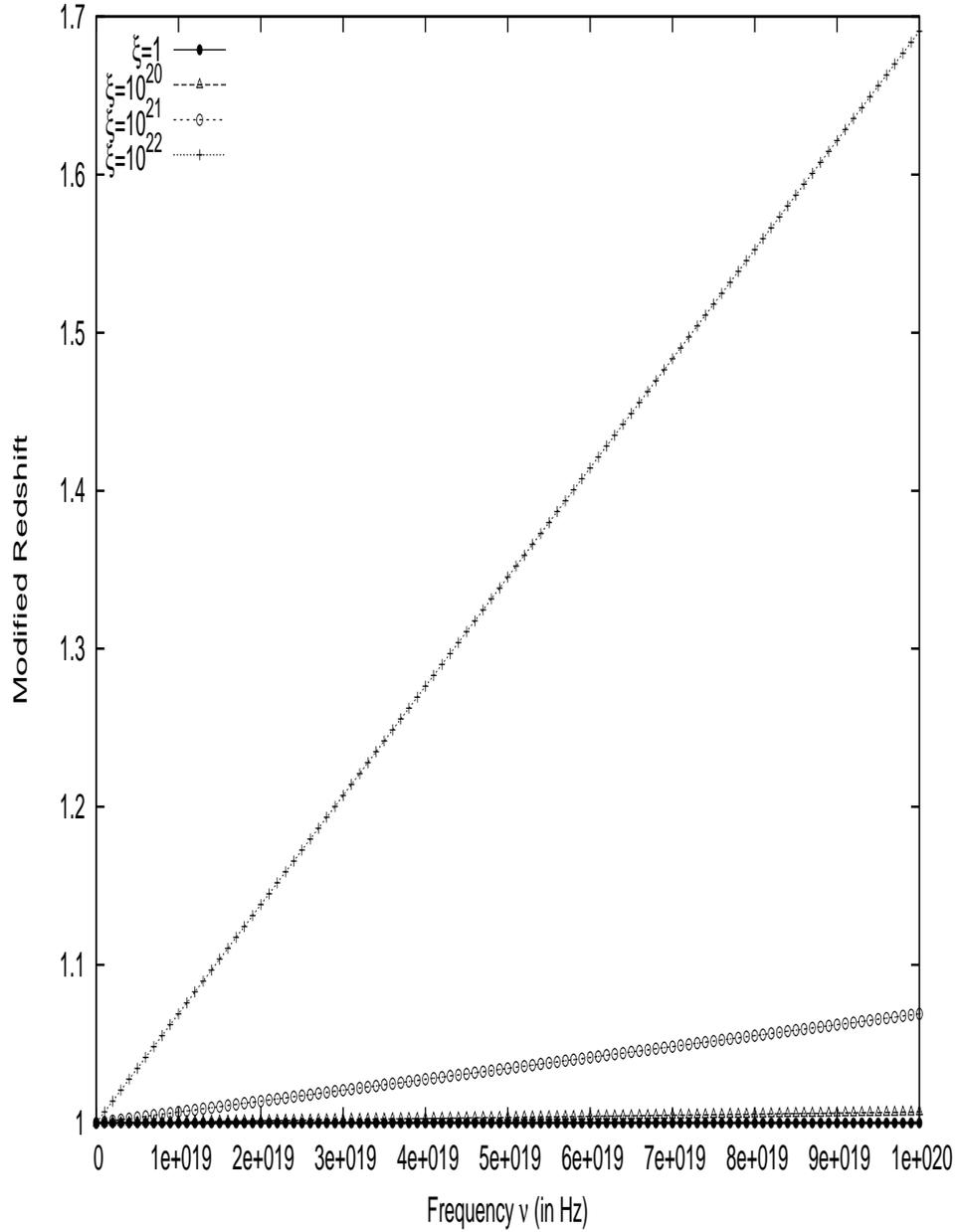}
 \caption[]{\label{fig1} Shows the variation of modified gravitational redshift ($\tilde{Z}$) as a function of  frequency ($\nu=\frac{\omega}{2\pi}$), at different values of rainbow function parameter ($\xi$)=1, $10^{20}$, $10^{21}$ and $10^{22}$, keeping fixed value of standard gravitational redshift (Z) equal to 1.}
\end{figure*}

In Fig. (\ref{fig1}),  a plot is made using the expressions (\ref{K-N4}), (\ref{Kerr}), (\ref{R-N}) and (\ref{Schwarzschild}), showing the variation of modified gravitational redshift ($\tilde{Z}$) as a function of  frequency ($\nu=\frac{\omega}{2\pi}$). This plot is made for the different values of rainbow function parameter ($\xi$)=1, $10^{20}$, $10^{21}$ and $10^{22}$, at fixed value of standard gravitational redshift (Z) equal to 1.
\section{Conclusions}\label{section5}
One can conclude from the present work that,
\begin{itemize}
  \item The general expression for the energy dependent gravitational redshift is derived for charged rotating body using the modified Kerr-Newman geometry.
  \item Using the appropriate boundary conditions in the obtained energy dependent gravitational redshift for charged rotating body in the modified Kerr-Newman geometry, one can obtain the corresponding gravitational redshift in various spacetime geometries such as Kerr geometry (rotating body), Reissner-Nordstrom geometry (static charged body) and Schwarzschild geometry (static body).
  \item One may obtain greater correction in the value of gravitational redshift, using the high energy photons.
   \item Knowing the value of gravitational redshift from a high energy sources such as Gamma-ray Bursters (GRB), one may obtain the idea of upper bounds on the dimensionless rainbow function parameter ($\xi$).  Also there may be a possibility to introduce a new physical scale of the order of  $\frac{\xi}{E_{Pl}}$.
\end{itemize}
\section{Acknowledgement}\label{section6}
We wish to thank Dr. Atri Deshmukhya, Department of Physics, Assam University, for very useful suggestions and inspiring discussions. A K Dubey is also thankful to Amritaksha Kar and Naznin Rahim Choudhury, Department of Physics, Assam University, for providing help and support in programming and plots.

\end{document}